\input{aipcheck}

\documentclass[
    ,final            
  ]
  {aipproc}

\layoutstyle{8x11double}

\begin{document}

\title{Performance of an X-ray single pixel TES microcalorimeter under DC and
  AC biasing}


\author{L.Gottardi}{
  address={SRON Netherlands Institute for Space Research,
  Sorbonnelaan 2, 3584CA, Utrecht, The Netherlands}
}
\author{J.van der Kuur}{
  address={SRON Netherlands Institute for Space Research,
 Sorbonnelaan 2, 3584CA, Utrecht, The Netherlands}
}
\author{P.A.J. de  Korte}{
  address={SRON Netherlands Institute for Space Research,
 Sorbonnelaan 2, 3584CA, Utrecht, The Netherlands}
}
\author{R. Den Hartog}{
  address={SRON Netherlands Institute for Space Research,
 Sorbonnelaan 2, 3584CA, Utrecht, The Netherlands}
}
\author{B. Dirks}{
  address={SRON Netherlands Institute for Space Research,
 Sorbonnelaan 2, 3584CA, Utrecht, The Netherlands}
}
\author{M. Popescu}{
  address={SRON Netherlands Institute for Space Research,
 Sorbonnelaan 2, 3584CA, Utrecht, The Netherlands}
}
\author {H.F.C.Hoevers}{
  address={SRON Netherlands Institute for Space Research,
 Sorbonnelaan 2, 3584CA, Utrecht, The Netherlands}
}
\author {M.Bruijn}{
  address={SRON Netherlands Institute for Space Research,
 Sorbonnelaan 2, 3584CA, Utrecht, The Netherlands}
}
\author{M.Parra Borderias}{
  address={Instituto de Ciencia de Materiales de Aragon
CSIC - Universidad de Zaragoza Facultad de Ciencias
50009 Zaragoza SPAIN}
}

\author{Y.Takei}{
  address={ISAS-Jaxa, Japan}
}


\begin{abstract}
 We are developing Frequency Domain Multiplexing (FDM) for the read-out
of TES  imaging microcalorimeter arrays for future  X-ray missions like
IXO.  In the FDM configuration the TES  is AC  voltage biased at  a well
defined frequencies (between 0.3 to 10MHz) and acts as an AM
modulating element. In this paper we will present a full comparison of
the performance of  a TES microcalorimeter under DC bias  
and AC bias at a frequency of 370kHz. In both cases  we measured
the current-to-voltage characteristics, the  complex
impedance, the noise, the X-ray responsivity, and energy resolution.
 The behaviour is very similar in both cases, but deviations in performances
 are observed for detector working points low
 in the superconducting transition ($R/R_{N}<0.5$). 
The measured energy resolution at 5.89keV is 2.7eV for DC bias and 3.7eV
for AC bias, while the baseline resolution is 2.8eV and 3.3eV, respectively.

\end{abstract}

\maketitle


\section{Introduction}
We are developing an imaging array of
 Transition Edge Sensor (TES) microcalorimeters and bolometers
for future X-ray and Infrared astronomy mission like IXO
 (International X-ray Observatory) and SPICA.
The experiment described here is part of the European-Japanese project
EURECA, which aims to demonstrate technological readiness of a 5 x 5
pixel array of TES-based micro-calorimeters read-out by two
SQUID-amplifier channels using frequency-domain-multiplexing (FDM) \cite{PietLTD12}. 
FDM requires amplitude modulation (AM) of the TES
signal and it can be achieved by biasing the TES with an AC voltage
bias source in a LC resonant circuit.
It is essential to demonstrate that the observed good performance of a single
pixels under constant voltage bias are maintained even when the TES
works as a modulator.
Several AC bias experiments with TESa have been performed so far with
micro-calorimeter \cite{Jan2004} and bolometers \cite{Lee2006}.
 In this paper we will present a full comparison of
the performance of  a  high resolution X-ray TES microcalorimeter under DC bias  
and AC bias at a frequency of 370kHz. 

\section{Experimental details}
The detector has been
integrated in a Janis two-stage ADR cooler
precooled at 3.5 K by a mechanical Cryomech Pulse Tube (PT). A
description of the measurement set-up can be found in
\cite{GottLTD12}.
The sensor is  a single-pixel of a 5x5 array and consists of a central Cu
absorber on top of a $Ti/Au$ TES deposited on a $Si_xN_y$ membrane,
which  provides a weak thermal link to the ADR bath temperature.
The transition temperature of the TES is $T_{C}\sim 100mK$, and the
normal state resistance is $R_{N}=143m\Omega$. All the measurements
presented here have been taken at a bath temperature of T=73mK and
zero magnetic field perpendicular to the TES. A full
characterization of this pixel under DC bias has been previously
described in\cite{YohLTD12,GottLTD12}. 
 
The current through the TES was
measured by a 100-SQUID array \cite{NIST} 
operated in flux locked
loop (FLL). The inductance of the SQUID input coil is $L_s\sim 70$ nH.
The SQUID array is direct read out by a low noise commercial
PTB-Magnicon electronics \cite{Magnicon}
in
the DC bias case and by a FrontEndElectronics(FEE) developed at SRON
in the AC bias case.

\begin{figure}[h]
\includegraphics[width=0.4\textwidth]{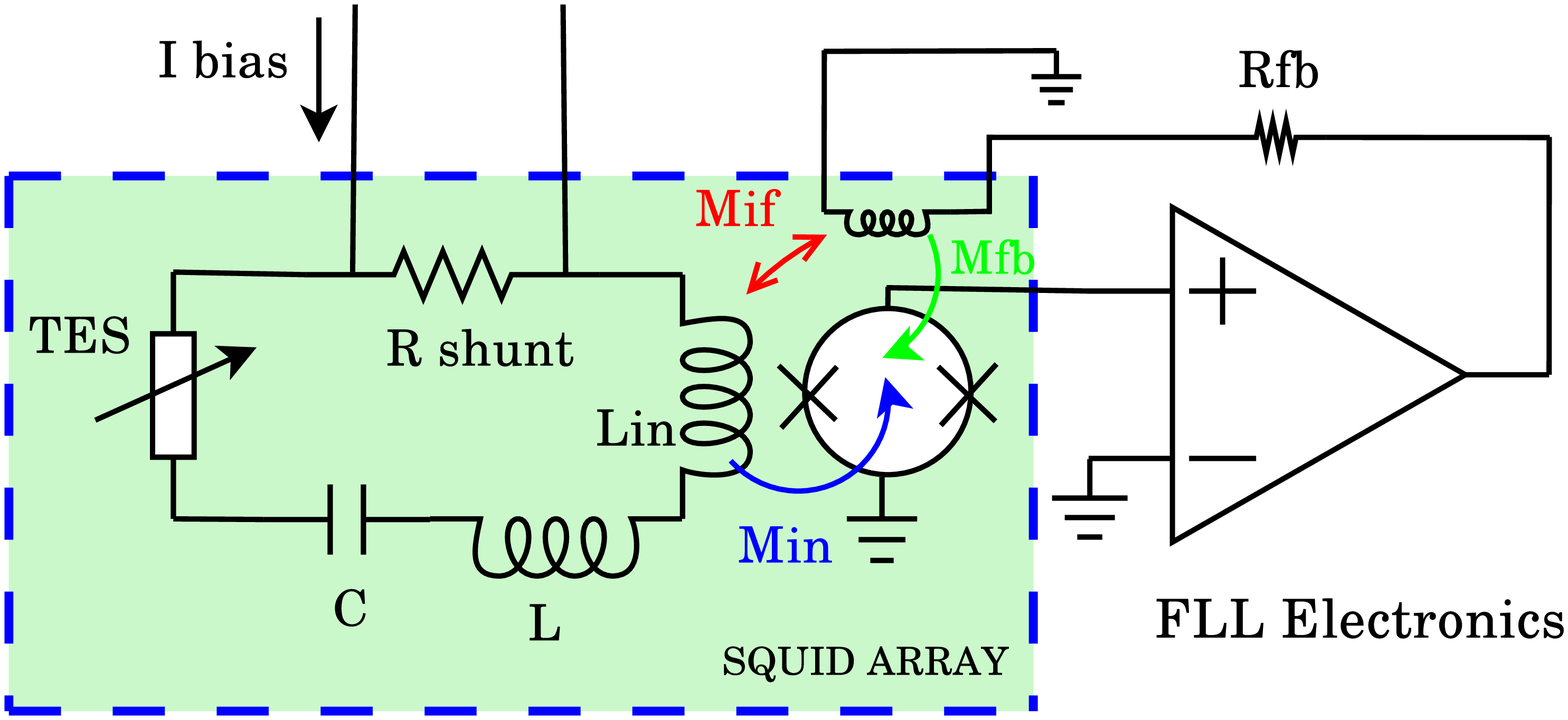}
\caption{Schematic drawing of AC bias and read-out circuit. \label{ACbiasreadout}}
\end{figure}

The scheme of the AC bias circuit is shown in Fig.~\ref{ACbiasreadout}.

The TES is
voltage biased with an alternating bias current (AC) through a shunt resistance. The resistance change of the
TES, induced by the detection of a photon, amplitude modulates the AC
current flowing through the TES. Under AC voltage bias $V(t)$, the Joule
power term ($V^2sin^2(\omega_0 t)/R_o$) switches on and off
periodically with twice the biasing frequency $f_0=\omega_0/2\pi$. 
 The capacitor C and the total
 $L_{tot}=L_{in}+L$ define the bias circuit resonant frequency
 $f_0=1/(2\pi\sqrt{L_{tot}C})$. In the experiment described here we
 use discrete components for L and C: L is a 700nH coil made of a Nb
 wire wounded around a Teflon cylinder and C is an NP0 capacitor with
 nominal value C=220nF. The SQUID read-out  amplifier is operated in
 a classical flux locked loop (FLL).


\section{Results}

\subsection{Current-voltage characteristic}
\begin{figure}[h]
\includegraphics[width=0.32\textheight]{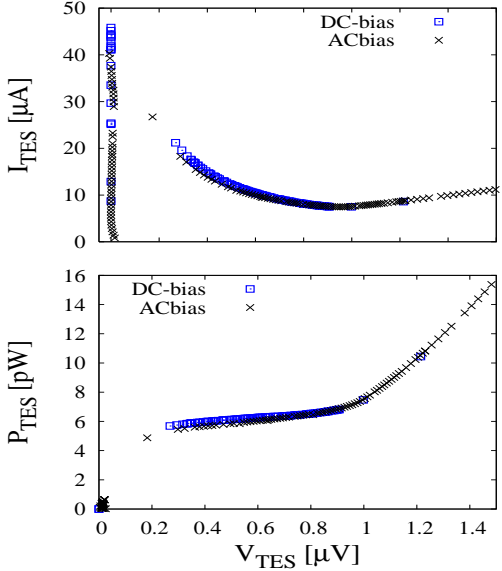}
\caption{Current and power to voltage characteristics under DC and AC bias.  \label{IVPV}}
\end{figure}

The Fig~\ref{IVPV} shows the current-to-voltage (I-V) and the
power-to-voltage (P-V) characteristics
of the microcalorimeter for both DC and AC bias. 

\begin{figure}[h]
\includegraphics[width=0.31\textwidth ,angle=270]{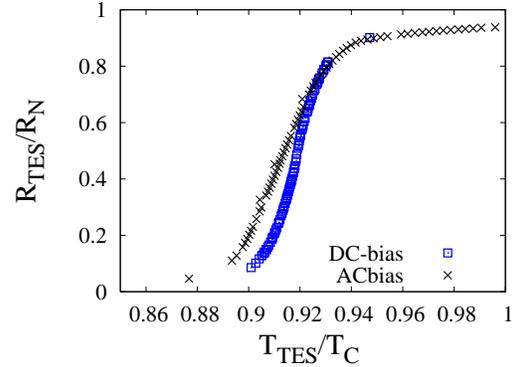}
\caption{Normalised resistance $R/R_N$ versus normalised temperature
$T/T_C$ for the DC and AC bias TES.  \label{RT}}
\end{figure}

The curves overlaps
for high TES bias voltage. Low in the transition a small deviation
between the curves is observed. This effect may be  due to the
non-linear behaviour of the TES resistance when the detector is AC
biased, since the resistance  is a function of
the sinusoidal bias current \cite{Jan2004}. This effect is difficult to
quantify and model because it depends on the dynamics of the
superconductor in the transition and it is very likely geometry
dependent. It may also be the result of a switching mechanism between
two or more current paths in the detector occurring at every cycle of
the carrier signal.       
Fig~\ref{RT} shows the normalised resistance versus the
normalised temperature curve
derived from the I-V characteristics. The transition is clearly
broadened in the AC bias case. For the same bias temperature lower in
the transition the
effective TES resistance under AC bias is about a factor of 2 higher.        

\subsection{Impedance measurements}
\begin{figure}[h]
\includegraphics[width=0.88\textwidth]{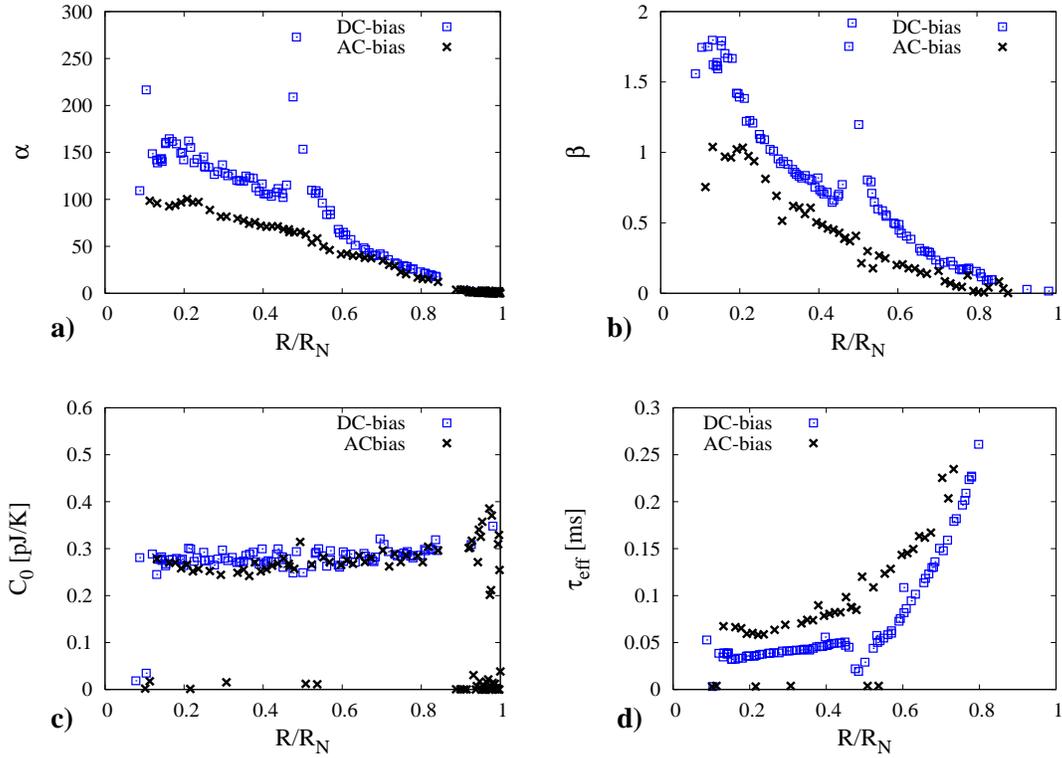}
\caption{The
parameters $\alpha,\beta$, the total heat capacity
$C=C_{TES}+C_{ABS}$ and the detector effective time constant are
plotted, in {\bf a},{\bf b},{\bf c} and {\bf d}, as a function of the bias point for both the DC and AC bias case. \label{Z}}
\end{figure}

From the  measurements of the  TES impedance $Z_{TES}(\omega)$  we can
derived the parameters, which  characterise the thermal and electrical
response  of the  TES \cite{Lindeman}.  The experimental  data  can be
fitted using  a second  order system involving  the TES  plus absorber
heat  capacity  and  a  dangling  heat capacitance  connected  to  the
TES-Absorber system via a thermal conductance $G_{A}$, as described in
\cite{YohLTD12}.    The   parameters   $\alpha=\partial   lnR/\partial
lnT$,$\beta=\partial  lnR/\partial  lnI$,   the  total  heat  capacity
$C=C_{TES}+C_{ABS}$  and  the  detector  effective time  constant  are
plotted in Fig~\ref{Z} as a function of the normalised TES resistance.

The major  difference between the AC  and DC bias case  is observed in
the $\alpha$ and $\beta$  parameters describing the sensitivity of the
TES   resistance   on   the   temperature  ($\alpha$)   and   on   the
current($\beta$) respectively.  Under DC bias both  the parameters are
larger, specially  lower in the transition,  and show a  peak at about
$R/R_{N}=0.5$. 
Features  like this are  often observed in  our pixels:
they are reproducible, but magnetic field and pixel dependent.  
They may  be caused by the inhomogeneous  current distribution in the TES.  
Under AC  bias this effect could be smoothed out by the change of the current direction during a carrier cycle. 
  
 The value of $\alpha$ and $\beta$
derived from $Z_{TES}(\omega)$ is consistent with the value
$\alpha_{IV}=\alpha/(1+\beta/2)$ derived from the R-T characteristic obtained
from the IV curve (Fig.~\ref{RT}). 

\subsection{Noise analysis}

\begin{figure}[h]
\includegraphics[width=1.0\textwidth]{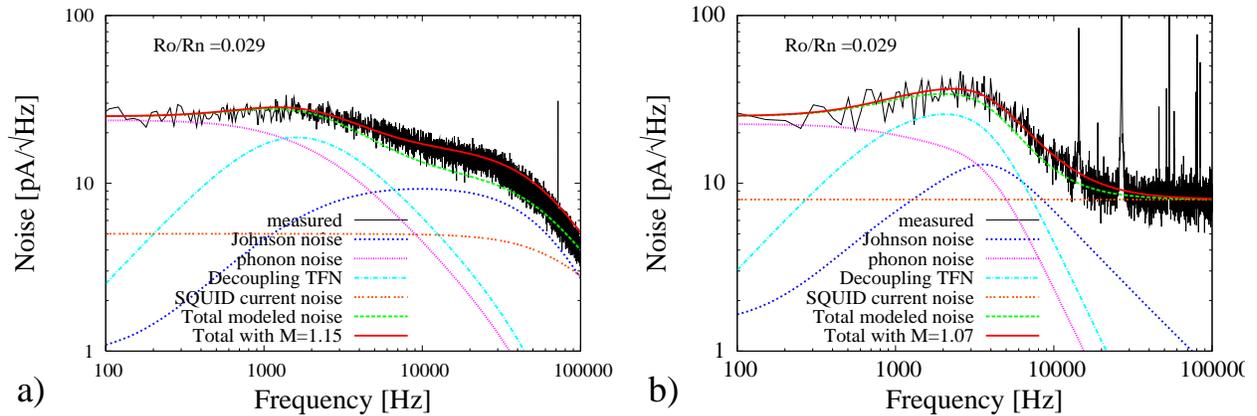}
\caption{Noise spectra at $R/R_N=0.29$ for a TES under DC bias (left)
and AC bias (right). In both case the model explained well the
observed noise. At this bias point a small excess noise  is observed at
high frequency with $M~1$ in both cases. \label{Noise}}
\end{figure}

One can use the parameters obtained  from the complex impedance to
model the detector noise. In Fig~\ref{Noise} the noise spectra
at the operating point $R/R_{N}=0.29$ is shown for the TES under DC
bias (left) and AC bias (right). The results from the models are
over-plotted.    The model noise contributions are:phonon
noise, TES Johnson noise  thermal
fluctuation noise (TFN) between TES-Absorber and dangling heat
capacity. Those noise sources describe very well the noise observed at
frequency below 5kHz both in the DC and AC bias configuration. In the
frequency range where the Johnson noise is mainly dominant a  non-modelled noise is present in the device. A useful
parameterisation of the unexplained noise is done by defining it in
terms of {\it M}-times the Johnson noise, like in Ullom et
al.\cite{Ullom}. Both the DC and AC bias  we observe $M<1$
at $R/R_N>0.5$ and $M> 1$ at bias point lower in the transition. 

\subsection{X-ray energy resolution}

We measured the energy resolution of the microcalorimeter by acquiring
the energy histogram of the $K\alpha$ complex at 5.89keV of a
$Fe^{55}$ X-ray source.
Due to the relatively small heat capacity of the pixel under
study the device does not operate in the small signal regime when a 6keV
photon is absorbed. The best energy
resolution is generally observed when the microcalorimeter is at
$R/R_N \sim 0.2$. Low in the transition the detector is faster and has
an effective time constant of about 100$\mu$s.  


Under DC an
X-ray energy resolution of $2.8\pm0.2$eV and a baseline resolution of
$2.7\pm 0.2$eV at the optimum working point $R/R_N=0.17$. Under AC
bias the best observed X-ray energy resolution is of $3.7\pm
0.3$eV. Nominal value of the X-ray energy resolution measured under AC bias
are however worse and fluctuates between 5eV and 6eV.
The baseline resolution is generally always better than the
X-ray resolution and is equal to $3.3\pm 0.2$eV. 
Due to the large inductance used in the circuit the lowest
point in the transition, achievable before oscillation in the electro-thermal feedback occur, is $R/R_N=0.3$. This results in a detector which is
slower and not optimally biased. 

The performance of the TES pixel with a central absorber described in
this paper is in general strongly dependent on the bias position. This
is true both in the DC and AC bias conditions. Pixels with similar
geometry have shown in the past bi-stable behaviour under DC bias and
even stronger dependence of the energy resolution to the working point.
We suspect that the broadening of the X-ray spectrum observed under AC
bias may be caused by instability in the sensor itself:  multiple current paths available in the TES, for example, could  induce the
sensor to change the bias point at any carrier cycle. 

Other possible sources of instability external to the TES  are: the
carrier amplitude, the read out gain, the loop-gain in the SQUID FLL.
We checked their stability in
different configurations of FLL electronics, SQUIDs and carrier
generators. Their contributions have been estimated to be smaller than 2eV.

\section{Conclusion}

We perform a full comparison of the performance of a single pixel TES
microcalorimeter under DC and AC bias ($f_{0}=370kHz$), respectively. 
In both cases  we measured
the current-to-voltage characteristics, the  complex
impedance, the noise, the X-ray responsivity, and energy resolution.

The behaviour of
the detector under AC bias begins to differ form the DC bias case at
working point lower in the transition. We remark that the calibration
of IV curves and complex impedance data taken under DC bias are very
sensitive to offsets around zero current.   A better analysis,
including the error estimation on  the parameters obtained from
the I-V characteristics and complex impedance data is needed to
guarantee a fair comparison.
 
Under DC an
X-ray energy resolution of $2.8\pm0.2$eV and a baseline resolution of
$2.7\pm 0.2$eV at the optimum working point $R/R_N=0.17$. Under AC
bias the best observed X-ray energy resolution is of $3.7\pm
0.3$eV. Nominal value of the X-ray energy resolution measured under AC bias
are however worse and fluctuates between 5eV and 6eV.
The baseline resolution is generally always better than the
X-ray resolution and is equal to $3.3\pm 0.2$eV.

In the future we are planning AC bias experiments using pixels with
different geometry and at different resonant frequencies to better
understand the behaviour of the detector as AM modulator.


\begin{theacknowledgments}
This work is financially supported by the Dutch Organisation for
Scientific Research (NWO) and ESA's Technical Research Programme contract 5417.
MPB would like to thank FPU program for predoctoral financial support.
\end{theacknowledgments}



\bibliographystyle{aipproc}   




\end{document}